\documentclass[a4paper]{article}

\usepackage{INTERSPEECH2020}
\usepackage{tabularx}
\title{An Overview on Audio, Signal, Speech, \& Language Processing for COVID-19}
\name{Gauri Deshpande$^{1,2}$, Bj\"orn W.\ Schuller$^{2,3}$}
\address{
  $^1$TCS Research India, India\\
  $^2$Chair of Embedded Intelligence for Health Care and Wellbeing, University of Augsburg, Germany\\
  $^3$GLAM -- Group on Language, Audio, \& Music, Imperial College London, UK}
\email{gauri1.d@tcs.com, schuller@ieee.org}

\begin{document}

\maketitle

\begin{abstract}
Recently, there has been an increased attention towards innovating, enhancing, building, and deploying applications of speech signal processing for providing assistance and relief to human mankind from the Coronavirus (COVID-19) pandemic. Many 'AI with speech' initiatives are taken to combat with the present situation and also to create a safe and secure environment for the future. This paper summarises all these efforts taken by the research community towards helping the individuals and the society in the fight against COVID-19 over the past 3-4 months using speech signal processing. We also summarise the deep techniques used in this direction to come up with capable solutions in a short span of time. This paper further gives an overview of the contributions from non-speech modalities that may complement or serve as inspiration for audio and speech analysis. In addition, we discuss our observations with respect to solution usability, challenges, and the significant technology achievements.

\end{abstract}
\noindent\textbf{Index Terms}: COVID-19, digital health, audio processing, computational paralinguistics, affective computing 

\section{Introduction}
As of today, there are more than 4 million confirmed cases of coronavirus-induced COVID-19 infection cases in more than 200 countries across the world. This led to increased need for screening, diagnostics, awareness, and post-care equipment and methods for a huge population. This not only demands for reaching out to a larger population in shorter time, but also to come up with effective solutions at each stage of the pathological, psychological, and behavioural problems. For the diagnosis purpose, the world health organisation (WHO) \footnote{www.who.int} has described the COVID-19 key symptoms as high temperature, coughing, and breathing difficulties, and an expanded list of symptoms that includes chills, repeated shaking with chills, muscle pain, headache, sore throat, and a new loss of taste or smell. These expanded list of symptoms are found to appear after 2-14 days of the onset of virus. However, the Coronavirus is so new to the mankind, that the unique symptoms of the virus and human immunity towards it are still found to vary from person to person. There are some cases found where a non-infected individual with no symptoms can still be a carrier of the disease.

Due to such uncertainties, it is advisable not to completely depend on digital sensing to diagnose the virus, where many physicians are also concerned about the false alarms that AI based diagnosing would generate. However, looking at the huge number of infected individuals across the globe, detecting early indications or screening will help in reducing this count. Continuous self-monitoring by the suspicious individuals can also help in decelerating and eventually stopping the spread of the virus. Not only does identification and monitoring need digital assistance, but also the post trauma phase would need it.

As per the WHO department of mental health and substance abuse, the current scenario of COVID-19 is susceptible to elevate the rates of stress or anxiety among individuals. Especially, lock-down and social distancing, quarantine and its after effects in the society might have adverse effects on the levels of loneliness, depression, self-harm, or suicidal behaviour. A special attention is needed for the health care providers, having to face the trauma directly and spending long working hours in such scenarios. Both physical and mental health needs have increased and require AI to provide faster and easy to access solutions.

\begin{figure}[t!]
  \centering
  \includegraphics[width=0.7\linewidth]{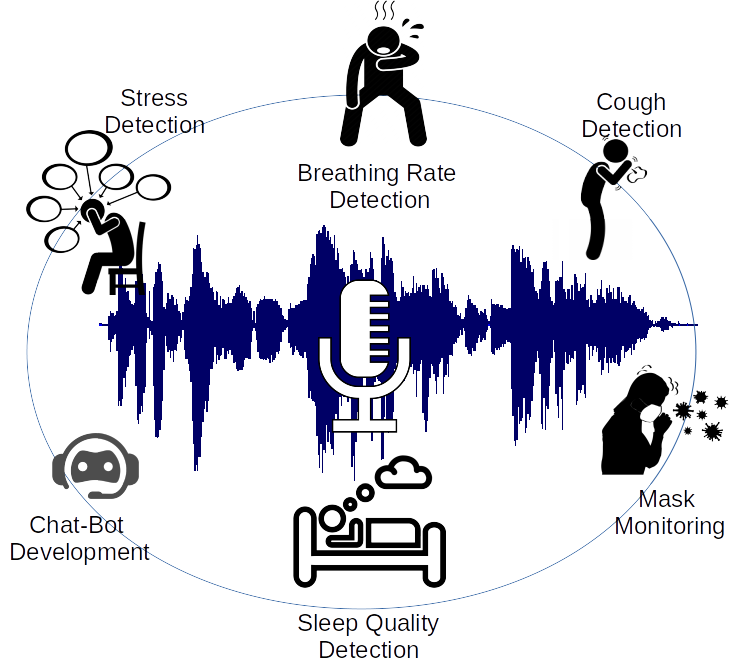}
  \caption{Speech and Audio Applications for COVID-19}
  \label{fig:speechApp}
\end{figure}

The authors of \cite{schuller2020covid} have already discussed about the opportunities, possible use cases, and the challenges that remain in the space of speech and sound analysis by artificial intelligence to help in this scenario of COVID-19. After this analysis was done, much more work has progressed in this direction which we are summarising in this paper. 
As depicted in Figure \ref{fig:speechApp}, we are giving an overview of the speech based applications for COVID-19 in this paper. Table \ref{tab:prevWork} explains past speech related work done to provide solutions for COVID-19 related health problems. This can be used as a ready reference by those who want to provide immediate solutions in this space. This table also describes the machine learning and deep learning technologies used on the given data-sets to provide mentioned accuracy. For each speech or audio application, there is a vast space in the literature. We have selected only 
a 
handful of them, considering their relevance in the COVID-19 situation. 
Not everything that can be developed, can be used in this scenario considering other factors such as social distancing and personal and environmental hygiene. Hence, the developments are required to be driven by guidance from the clinicians and health care providers. 

\begin{table*}[th]
  \caption{Past speech and audio analysis related to COVID-19 health problems}
  \label{tab:prevWork}
  \centering
  \begin{tabularx}{\linewidth}{X | l | X | X | X }
    \toprule
     \textbf{Application}  & \textbf{Ref.}      & \textbf{Technology}  & \textbf{Dataset}  & \textbf{Accuracy}\\
    \midrule
    Tuberculosis cough detection & \cite{miranda2019comparative} & STFT, MFCC, MFB with CNN & Google audio set extracts from 1.8 million Youtube videos and Freesound audio database \cite{van2016vu} & .946 AUC \\
    
    Asthmatic detection & \cite{yadav2020analysis} & INTERSPEECH 2013 Computational Paralinguistics Challenge baseline acoustic features \cite{schuller2013interspeech} & Speech from 47 asthmatic and 48 healthy controls & 78\,\% Accuracy   \\
    
    Avoiding speech recording while sensing cough and hence providing privacy attribute & \cite{larson2011accurate} & PCA on audio spectrograms, FFT coefficients and RandomForest Classifier & Acted cough from 17 patients having cough due to common cold (n=8), asthma (n=3), allergies (n=1), and chronic cough (n=5) & True positive rate of 92\,\% and false positive rate of 0.5\,\%  \\

    Obstructive sleep apnea (OSA) detection from breathing sounds during speech & \cite{simply2018obstructive} & MFCCs with single layer neural network for breathing detection and MFCCs, energy, pitch, kurtosis and ZCR with SVM for OSA classification & 90 Male subjects' speech and sleep quality measures using WatchPAT \cite{gan2017validation} & Cohen's kappa coefficient of 0.5 for breathing detection and 0.54 for OSA detection \\
    
    Automatic speech breathing rate measurement & \cite{routray2019automatic} & Cepstrogram and Support Vector Machine with radial basis function & Speech recordings of 16 participants of age group 21 years mean & 89\,\% F1 score and RMSE of 4.5 breaths/min for the speech-breathing rate \\
    
    Breathing signal detection for conversational speech & \cite{nallanthighal2019deep} & Spectrogram with CNN and RNN & 20 healthy subjects' speech recorded using microphone and breathing signal using two respiratory transducer belts & 91.2\,\% sensitivity for breath event detection and mean absolute error of 1.01 breaths per minute for breathing rate estimation \\
    
    Stress detection from speech in situations such as car accident, domestic violence, situations close to death, and so on & \cite{partila2019human} & LLD and functional features extracted using openSMILE \cite{eyben2010opensmile} with k-NN, SVM, and CNN classifiers & 31 emergency call recordings of the Integrated Rescue System of 112 emergency line from Czech Republic & Accuracy of 87.9\,\% with SVM and 87.5 \% with CNN in classifying stress from neutral speech \\
    \bottomrule
  \end{tabularx}
\end{table*}

\section{Screening and Diagnosing for COVID-19}

There exists a fine line between screening and diagnosing, where screening gives an early indication of the presence of a disease and diagnosing confirms the presence/absence of disease. Screening is probabilistic, whereas diagnosis is binary in nature. We will talk about different algorithms/applications using audio processing for screening and diagnosis of COVID-19.

\subsection{Cough Detection}
\label{cough}

Cough detection is not only identifying the cough sound and differentiate it from other sounds such as speech and laughter but also, to identify COVID-19 specific cough. This requires to collect the COVID and non-COVID cough sound samples so as to develop an AI model that can differentiate between them on its own. 
For development of such a model, relevant data needs to be collected. Cough audio samples can be collected using a simple smartphone mic, hence, major efforts are seen in collecting it
using smartphone mics such as done by the Carnegie Mellon University \footnote{https://cvd.lti.cmu.edu/}
and the Cambridge University \footnote{https://www.covid-19-sounds.org/en/}. 
Both the universities have provided platforms for the general population to upload their cough sounds along with some additional information such as their age, gender, location, and if they had been tested positive. They would need these audio samples to build their machine learning algorithms for identifying COVID-19 cough. The Cambridge University's web based platform asks the participants to also read a sentence so that their speech can also be recorded. Hence, their app considers the sounds of cough, breathing and voice. Coughvid \footnote{https://coughvid.epfl.ch/} is another such app from EPFL (Ecole Polytechnique Fédérale de Lausanne) to detect COVID-19 cough from other cough categories such as normal cold and seasonal allergies. 'Breath for Science' \footnote{https://www.breatheforscience.com/} -- a team of scientists from NYU --, have developed a web based portal to register the participants where they can enter similar details along with phone number. On pressing a 'call me' button, the participants receive a callback where they have to cough three times after the beep. With this, they plan to create a dataset of cough sounds for the research purpose. 
Another web interface 'CoughAgainstCovid' \footnote{https://www.coughagainstcovid.org/} for COVID-19 cough sample collection is an initiative by Wadhwani AI group in collaboration with Stanford University \footnote{https://www.stanford.edu/}. 

There are others who are ready with COVID-19 cough identifiers such as a smartphone app described in \cite{imran2020ai4covid}, which detects the COVID-19 cough.

Although these algorithms are attaining high accuracy levels of above 90\,\%, however, from the hygiene perspective it is not advised to cough on open surface. As explained in \cite{singhal2020review}, the infection is primarily transmitted through large droplets generated during coughing and sneezing by symptomatic and also by asymptomatic people before onset of symptoms. These infected droplets can spread 1-2\,m, deposit on surfaces and can remain viable for days in favourable atmospheric conditions but can be destroyed in a minute by common disinfectants. Hence, for the collection procedures, it is important to remind the participants to cover the mouth and then only provide the cough sound samples.  Otherwise, this may result in further spreading of the disease. Also, after giving the samples, the smartphone surface should be applied with an disinfectant. 

\subsection{Breathing Analysis}
\label{breathing}
As discussed before, shortness of breath is also one of the symptoms of the virus for which, the smartphone apps are designed to capture breathing patterns by recording their speech signal. Breathing pattern detection has found applications in spectrometry to analyse lung functionalities as well. The authors of \cite{nallanthighal2020speech} have attempted to correlate high quality speech signals captured using an Earthworks microphone M23 at 48\,kHz with the breathing signal captured using two NeXus respiratory inductance plethysmography belts over the ribcage and abdomen to measure the changes in the cross-sectional area of the  ribcage and abdomen at the sample rate of 2\,kHz. They have achieved a correlation of 0.42 to the actual breathing signal and a breathing error rate of 5.6\,\% and sensitivity of 0.88 for breath event detection. In the Breathing Sub-challenge of Interspeech 2020 ComParE \cite{schuller2020interspeech}, the authors have achieved a baseline pearson's correlation of 0.507 on a development, and 0.731 on the test dataset, respectively. They have used two piezoelectric respiratory belts for capturing breathing patterns. In addition to the speech signals, blow sound can also help in performing spirometry tests. The authors of \cite{trivedy2020design} have developed a smartphone based automatic disease classification application built around the spirometry tests. 


\subsection{Chat-Bots}
\label{chatbot-S}
As the count of positive COVID-19 cases are increasing every day, it brings up the need of automating the conversation a physiologist would have to understand the presence of symptoms in an individual. 
Microsoft and CDC have come up with a chatbot named ``Clara'' for initial screening of COVID-19 patients by asking them questions and capturing the responses. This uses speech recognition and speech synthesis technologies. The risk-assessment test is designed based on advice from the WHO and the Union Ministry of Health and Family Welfare India \footnote{https://www.mohfw.gov.in/}. ``Siri'' from Apple is also updated to answer the variations of the general question, 
``Siri, do I have the coronavirus?'' based on WHO guidelines. If a person shows severe symptoms, then it is advised by Siri to call 911. Similarly, Alexa is also updated with answering COVID-19 screening question based on CDC guidelines.
Considering the trauma that the health care providers are going through, a web-based chat-bot named Symptoma developed in \cite{martin2020artificial}, is a significant step. It can differentiate 20\,000 diseases with an accuracy of more than 90\,\% and can identify COVID-19 from other diseases having similar symptoms with an accuracy of 96.32\,\%.

\subsection{Chest X-Ray}

Multiple efforts by several groups are put in the direction of developing a classifier to detect COVID-19 symptoms using chest X-Ray, such as that of in \cite{afshar2020covid}, \cite{apostolopoulos2020covid}, and many more. However, as stated in \cite{rubin2020role}, a multinational consensus is that Imaging is indicated only for patients with worsening respiratory status. Hence, it is advised for only those patients who have moderate-severe clinical features and a high pre-test probability of disease. Hence, unlike Sections \ref{cough}, \ref{breathing}, and \ref{chatbot-S}, such measures are not suggestive for early identification and are preferred for diagnosis in a clinical setup. 

As concluded by \cite{wong2020frequency}, the chest X-ray findings in COVID-19 patients were found to be peaked at 10-12 days from symptom onset. Also, it is still required to visit a well-equipped clinical facility for such approaches. On a positive note, in the presence of mobile X-Ray machines, this approach can help in speeding up the diagnosis. The authors of \cite{li2020coronavirus} have found from an experimental outcome that the chest X-Ray may be useful as a standard method for the rapid diagnosis of COVID-19 to optimise the management of patients. However, CT is still limited for identifying specific viruses and distinguishing between viruses.

\section{Monitoring}

The precautions for stopping the spread of the virus include social distancing, wearing a mask, and maintaining respiratory hygiene. Especially at public places, the concerned authorities can monitor the population to confirm that the precautionary measures are adopted by them. This section describes such monitoring tools developed for the COVID-19 scenario.

\subsection{Face Mask Detection from Voice}

As described in Section \ref{cough}, collecting cough samples without covering mouth can lead to further spreading of the disease, mask wearing has to be mandated for donating a cough sample which requires an app to detect if the donor is wearing a mask or not. This year's Interspeech 2020 ComParE challenge \cite{schuller2020interspeech} features a mask detection sub-challenge, where the task is to recognise whether the speaker was recorded while wearing a facial mask or not. The results from this challenge will be useful to develop a voice monitoring tool which detects the mask or no-mask of a speaker. Also, these algorithms serve as a pre-step for other speech processing algorithms such as speech recognition, emotion recognition and cough detection.

\subsection{Cough and Breathing Analysis}

While at quarantine, the doctors need to monitor the cough history of patients, which can be done with a continuous cough monitoring device.
After we cross the crisis and organisations think of resuming the operations, continuous monitoring of common spaces such as canteens and lobbies can be realised to record the COVID specific coughs.
One such monitoring application is developed by the FluSense platform in \cite{al2020flusense}. It is a surveillance tool to detect influenza like illness from hospital waiting areas using cough sounds. 
Continuous monitoring and identification of abnormalities from the breathing rate has been done by \cite{wang2020abnormal} using image processing.

\subsection{Mental Health -- Emotion Detection}

The disease spread has equally affected the physical and mental health of the individuals, which is primarily due to plenty of mandatory precautionary measures such as social-distancing, work from home and the quarantine procedures which usually takes at-least 15 days for the patients to be alone. Also, the health care providers owing to their hectic routines followed by quarantine days are subject to undergo mental health issues. To cater for the growing need of addressing this issue, not only there is a high demand but the physical presence of the available psychologists is also missing. As found in \cite{patel2020rapid}, the COVID-19 pandemic has generated unprecedented demand for tele-health based clinical services. 

Among several initiatives taken against mental health issues such as stress, anxiety, and depression, we are yet to see these emotions being analysed from the speech signal during the COVID-19 period. This demands for the relevant data. Very recently, a study is conducted by the authors of \cite{han2020early} on the speech signal of COVID-19 diagnosed patients. The 
behavioural 
parameters detected from speech includes, sleep quality, fatigue, and anxiety considering corresponding self-reported measures as ground truth and have achieved an accuracy of 0.69. 
This year's ComParE challenge at Interspeech 2020 \cite{schuller2020interspeech} has an Elderly Emotion Detection sub-challenge, where the speech captured from elderly people has to be classified into Low, Medium, and High classes of Valence and Arousal. It is found that specific age groups such as that of 
elderly above 60 years are more prone to the infection, due to which this age group needs to follow the restrictions posed by the pandemic for a larger period in future as well. This shows that it will be crucial to understand the effects of this phase on their psychological parameters such as emotions.

\subsection{Face Mask Detection from Image}
One of the precaution 
measures 
while stepping outside home suggested by the WHO to reduce the chance of getting infected or spreading COVID-19 is to wear a mask. 
Also, the mathematical analysis presented in \cite{eikenberry2020mask} suggests that wearing a mask can reduce community transmission and can help reduction in the peak death rate. Hence, similar to what has been outlined above, it is important that the concerned authorities keep a check on people wearing masks or not at public places, especially where the population is quite dense such as airports, railway stations, and hospitals. 

The authors of \cite{wang2020masked} have provided the Masked Face Detection Dataset (MFDD), Real-world Masked Face Recognition Dataset (RMFRD), and the Simulated Masked Face Recognition Dataset (SMFRD) for the detection of masked and unmasked face using image processing techniques. They have achieved 95\,\% accuracy in recognising masked faces. An Apple store app developed by LeewayHertz can be integrated with the existing setup of an 
Internet Protocol (IP) camera for getting alerts on detecting no-mask on a face. The system provides a facility to add phone numbers for receiving the alerts and also mechanism to see the face not wearing a mask for the admin.

\subsection{Text Analysis}

The authors of \cite{guney2020using} have used machine learning techniques to categorise sentiments from the responses to Press Ganey patient experience surveys. From this analysis, it is found that the patients have expressed negative comments about the cleanliness and logistics and have given positive comments about their interactions with clinicians.
In \cite{ahmed2020dangerous}, the twitter data with 5GCoronavirus hashtag is analysed to understand the drivers of the 5G COVID-19 conspiracy theory and strategies to help in reducing the spread of mis-information circulating in society.

\section{Awareness}

Speech recognition and synthesis algorithms have been widely used in the development of chat-bots to provide human like interactions. In this time of crisis, chat-bots are helping in spreading the valuable information about COVID-19 to end users. Once an infected gets cured, they can help researchers with not only their experience but also with donating components such as plasma. CDC has been encouraging recovered people to donate their plasma for development of blood related therapies. For collection of plasma, Microsoft has developed the chat-bot \footnote{https://covig-19plasmaalliance.org/} which interacts with individuals to gather the required information such as, the duration for which they are tested negative, their age, weight, and also takes their pin-code to help them know their nearest donation center. 



\section{Next Steps and Challenges}

Since the work culture is moving more towards working-from-home, it will be required to detect some behavioural parameters such as fatigue and sleepiness for the self monitoring of the working professionals. 
The behavioural parameters such as stress or anxiety need greater attention to be paid in these days. 
A major challenge given the social distancing norms, is getting the relevant and accurate speech data for developing machine learning models. Speech enabled chat-bots can play a significant role in this. There are other challenges as well from the design perspective of chat-bots. The authors of \cite{miner2020chatbots} have expressed both positive effect and drawbacks associated with using chatbots for sharing the latest information, encouraging desired health impacting 
behaviours, 
and reducing the psychological damage caused by fear and isolation. This shows that the design of chat-bots should be well thought of for using them, otherwise, they might have negative impact as well. 
An optimistic approach in these difficult times has been to work towards safe and secure environment for the post pandemic situation so that the society gain the trust and confidence back. This shows the need of accurate and reliable screening and monitoring measures at public places. 

\section{Conclusion}
There is a vast space of physical and mental wellness which needs fast and usable solutions from the digital sensing space. 
Most of the speech based research 
works 
for screening and monitoring of COVID-19 are ready to be customised. Chat-bots, although having the largest possibilities of use-cases, are still in the need of context aware designs. 
As most of the first level symptoms are related to respiratory system, continuous monitoring of the patients using audio signal processing based techniques can assist the clinicians or health care providers in providing services while following social distancing norms. 
A multi-modal approach where all three modalities, audio and speech, text, and image together work towards creating a solution framework will be likely the most promising avenue of future efforts.

\section{Acknowledgements}
We would like to thank all researchers, health supporters, and others helping in this crisis. Our hearts are with those affected and their families and friends.  
We acknowledge funding from the German BMWi by ZIM grant No.\ 16KN069402 (KIrun).

\bibliographystyle{IEEEtran}

\bibliography{mybib}


\end{document}